\documentclass[draft]{agujournal2018}

\usepackage{apacite}
\usepackage{url} 

\draftfalse

\journalname{JGR-Space Physics}

\begin{document}

%
%


\title{Testing the Radiation Pattern of Meteor Radio Afterglow}

%
%




\authors{S. S. Varghese\affil{1}, K. S. Obenberger\affil{2}, G. B. Taylor\affil{1}, J. Dowell \affil{1}}


\affiliation{1}{University of New Mexico, Albuquerque, NM, USA}
\affiliation{1}{Air Force Research Laboratory, Kirtland AFB, NM, USA}




\correspondingauthor{Savin Shynu Varghese}{savin@unm.edu}




\begin{keypoints}
\item We have detected 32 co-observed meteor radio afterglows (MRAs) and \newline
21 transmitter reflections from meteors (meteor scatter events) using \newline two stations of Long Wavelength Array (LWA).

\item Flux and distance of the co-observed events from each LWA station \\ were measured precisely to test the radiation pattern using flux-distance \\ relation for an isotropic emitter.

\item The MRAs appear to be distinct from the meteor scatter events \\ and are consistent with an isotropic radiation pattern.
\end{keypoints}

%
%

\justify
\begin{abstract}

Radio emission from meteors or meteor radio afterglows (MRAs) were first detected using the all-sky imaging capabilities of the first station of the Long Wavelength Array (LWA1). In this work, we use the recently commissioned LWA Sevilleta (LWA-SV) station along with the LWA1 to carry out co-ordinated observations. The combined all-sky observations with LWA1 and LWA-SV have co-observed 32  MRAs and 21 transmitter reflections from meteors (meteor scatter events) which are believed to be specular reflections from overdense trails.  The flux density of the events observed by each station were measured from the all-sky images. Triangulating the angular direction of events from each station gave the physical location and the distance of the event to each station. The luminosity of the events in each station were calculated using the flux distance relation for an isotropic source. The luminosity distribution for MRAs and meteor scatter events observed by each station shows a clear distinction between these two types of events as the ratio of luminosities are closer to unity for MRAs than the meteor scatter events. Furthermore, we find that MRAs follow an isotropic radiation pattern. This suggests, either a complete incoherent emission mechanism or an incoherent addition of coherently emitting small regions within the meteor trail.

\end{abstract}

%
%

%


%
%
%
%
\section{Introduction}

Meteors occur when high velocity, solid material collides with the Earth's atmosphere, ablates, and ionizes to form a plasma trail. The recent discovery of high frequency (HF; 3-30 MHz) and very high frequency (VHF; 30-300 MHz) radio emission from meteors \citep{ob14b} has opened a new window to study meteor physics. The all-sky imaging capabilities of the first station of the Long Wavelength Array (LWA1) was utilized to discover the previously unknown radio emission from meteors or meteor radio afterglows (MRAs). Combined radio observations with optical cameras have confirmed that the afterglow emission is produced by optical meteors and occurs within few seconds of the optical activity. The unpolarized MRA emission usually lasts between a few seconds to a few minutes with a characteristic light curve pattern of a fast rise and slow decay \citep{ob14b}.

Follow-up observations with the beamformed mode of the LWA1 have collected the dynamic spectra of 4 MRAs \citep{ob15b,ob16a}. The dynamic spectra  captured between 22 - 55 MHz suggest that the afterglow emission is broadband and the flux density follows a power law dependence with frequency,  having higher flux at lower frequencies \citep{ob15b,ob16a}.

The intrinsic MRA emission is presumed to be different from the well studied transmitter reflections from meteors (meteor scatter events) which we also detect in LWA stations \cite{ob14b}. Transmitter echoes reflecting off meteor trails are strongly polarized \citep{clo11} and are narrow band in frequency, whereas MRAs are largely unpolarized and are broadband in frequency.


Transmitter reflections from meteor trails have been well studied to understand meteor physics. In general, reflection events from meteors can be classified as specular, non-specular, and head echoes. The purpose of this paper is to study the radiation pattern of MRAs and to compare it with what we expect from meteor scatter events.

Historically, the most commonly observed reflection cases are specular and they satisfy the law of a regular reflection \citep{wis96}.  Specular echoes arises from the Fresnel scattering when the radar beam is pointed perpendicular to the meteor trail \citep{clo08}. Specular reflections from meteor trails are directional, and it is impossible for both stations to detect the same specular reflection. However, since the mesospheric neutral winds distort the shape of a meteor trail, it is possible for different regions of a single overdense meteor trail to produce a specular reflection towards both stations. In such cases the power observed by both stations should be largely independent of each other. Underdense trails usually last for less than a few tenths of a second and the neutral winds barely affect them \citep{wis96}. Therefore, it is unlikely to observe a specular reflection from an underdense trail simultaneously in both stations.

The radiation pattern from head echoes are far less directional than specular echoes \citep{clo04,mar15}. However, since they only last for the short period of time when the meteoroid is ablating, such echoes have little relevance to the study of MRAs, which can last for up to minutes after ablation \cite{ob14b}.

Non-specular reflections, also known as RSTEs (Range Spread Trail echoes)\textit{,} have been studied for the last few decades \citep[and references therein]{mc49,ma04,chau14}. One type of non-specular reflection occurs when a radar beam is pointed perpendicular to the magnetic field as opposed to perpendicular to the meteor path.  \citet{he62} first described them as magnetic field-aligned irregularities. Here, field-aligned irregularities form within the meteor trail and allow for coherent scatter back to the radar. Presumably this could also allow for passive scattering where the geometry of the radiation source, magnetic field, and receiver satisfy the law of reflection. Like specular reflections, non-specular reflections from field-aligned irregularities are highly directional \citep{clo08}, and it is  unlikely that such events could be simultaneously observed in both stations.

In addition to field-alligned echoes, \citet{kel04} has suggested and \citet{chau14} has shown that non-specular radar echoes can arise from non-field-aligned irregularities due to turbulence induced by charged dust within the meteor trail. 
These irregularities may not be field aligned, but they can be aligned with respect to the meteor trail axis.  In that case, they will not scatter isotropically and will have a preferred direction just as specular trails. Therefore it is unlikely that both LWA1 and LWA-SV could see the same amount of power scattered by a non-field aligned non-specular meteor, and we note that in general, such echoes would be significantly dimmer than specular echoes \citep{chau14}. 

Comparing different known scattering mechanisms, it is difficult to imagine a broadband, unpolarized transmitter powerful enough to scatter off of meteor trails isotropically.
 We can therefore assume that if the radiation coming from MRAs is for the most part isotropic, then MRAs cannot be coming from any sort of scattering process, and must indeed be due to self-generated emission.

In this paper,  we present observations from 32 MRAs that were co-observed by both stations. These events are consistent with previous observations of MRAs, which are characterized by broad-band, unpolarized HF/VHF emission. We compare the power measurements of these dual station MRAs with 21 co-observed narrow-band, polarized echoes, which we assume to be specular echoes from meteor trails that have been warped  by the neutral winds, such that specular echoes are observed in both stations nearly simultaneously.

\section{Observations}
The combined observations of MRAs and meteor scatter events were carried out with the LWA1 and LWA-SV stations. LWA1 \citep{tay12} is the first station of the Long Wavelength Array radio telescope and is collocated with the Karl G. Jansky Very Large Array (VLA). The recently commissioned LWA-SV station is located 75 km North East of LWA1 on the Sevilleta National Wildlife Refuge \citep{cra17}. Both stations operate between 10 and 88 MHz frequency range and have a similar physical layout. The core of each station consists of 256 dual polarization dipole antennas arranged in the form of an ellipse 100 $\times$ 110 m across. In addition, 5 outrigger antennas in LWA1 and one in LWA-SV are located at 200-500m away from the center of the arrays. The MRAs and meteor scatter events were detected using the all-sky images produced by the LWA All-Sky Imager (LASI) which is the backend correlator. The correlator converts the collected raw voltages from each antenna to an image of the sky every 5 seconds. More details of LASI can be found in \citep{ob15a}. This work utilized 13771 hours of all-sky images collected at the two LWA stations between May 2016 and February 2019.

\section{Detection of co-observed MRAs and meteor scatter events}
The transient events from all-sky images are found using the transient search pipeline. The pipeline is based on an image subtraction process which subtracts the average of four previous images from a running image and pixels with flux greater than six sigma threshold above the image noise are marked as transients. The imaging and transient search pipeline is discussed in  \citep{ob14b,ob15a,var19}.  The transient candidate list from each station are compared to find the coordinated events which happened simultaneously in both stations. Depending on the angular direction in which the coordinated events are detected in each station, they are classified as MRA candidates and cosmic transient candidates. For example, a transient event (like a reflection from an airplane, lightning or local radio frequency interference) happening above LWA1 at an altitude of 5 km will not be detectable by LWA-SV as both stations do not share the same sky at lower elevations. As the elevation increases, the shared region of sky between the two stations increases. Both stations share 99\% of the sky at elevations greater than 1430 km. If a transient event, from it's physical location, makes an angle of three degrees along the 75 km baseline between two LWA stations, then this event corresponds to an elevation of 1430 km. Therefore, any event happening at great distances (such as solar system objects or cosmic sources) will have a difference between the angular coordinates measured by each station to be less than three degrees. For events like MRAs happening between 90-130 km, the events detected by the two stations will have angular differences between 30-45 degrees in their measured coordinates. Therefore, if the angular difference between coordinates of the event from each station is greater than 3 degrees, it is classified as an MRA candidate otherwise it is classified as a cosmic transient candidate. The process of finding co-observed events and the event classification is described in more detail in \citet{var19}.

The MRA candidates detected by each station can be MRAs, meteor scatter events or scintillations of radio sources caused by the ionosphere. Meteor scatter events are transmitter signals reflecting off the meteor plasma trails. A transmitter signal with frequency less than the plasma frequency of the meteor plasma trail can reflect off the trail to LWA stations. As the trail is warped by the neutral winds, a specular meteor can be observed by both stations simultaneously. The origin of meteor scatter event at each station can be from the same or different transmitters.
Scintillation is a propagation effect which occurs as the radio waves from cosmic sources pass through Earth's ionosphere. The turbulent magnetized plasma in the ionosphere cause refraction, scattering and interference of waves resulting the rapid fluctuation in the observed flux and changing the actual source position by few degrees. More details of dealing with scintillation and their statistics can be found in \citet{var19}. 
Our comparison method can correlate all events seen from both stations as transient events.  We find MRA - MRA, meteor scatter - meteor scatter , scintillation - scintillation and meteor scatter - scintillation event pairs from both stations. For the purpose of this project we are interested in finding the real pairs coming from the same meteor, MRA - MRA and meteor scatter - meteor scatter pairs. From the list of candidates, we need to remove the false positive events like scintillation - scintillation and meteor scatter - scintillation pairs.

Triangulation can be used as an effective tool to eliminate false positives. It is a process of forming a triangle from two known points and directions to find the third unknown location. The azimuth and altitude of the source measured from each station are used to find the latitude, longitude and elevation of the source. For a real event, the calculated coordinates of the source from each station should agree with each other. We filter to select events which have an elevation between 60-150 km and having less than 5\% error in the difference between the coordinates of the source calculated from each station.   This method can find the  physical location of a real transient observed in two stations with good accuracy. At the same time,  triangulation criteria fails for scintillation and meteor scatter happening at different physical locations on the sky, and they are marked as  false positives. This method works well, removing 95\% of the false positive events. However, it is possible to have false positive events observed from each station at particular angular directions which can satisfy the triangulation criteria.  Such events are rare and the detected events have to be analyzed later by examining their light curves.  The process of finding transients in each station, their comparison, classification and triangulation of MRA candidates are automated.

Further confirmation of event characteristics requires the analysis of Stokes I, Q, U and V light curves of the events from each station.  \citet{ob14b} clearly shows that MRAs have little to no linear or circular polarization, whereas meteor scatter events tend to be highly polarized. While identifying MRAs, we chose all events which had a linear or circular polarization fraction less than 30\% and the events with high amount of polarization were classified as meteor scatter events.

While the polarization test appears to be a reliable method for separating MRAs from scatter events, MRAs can be further differentiated by their light curves, which are typically characterized by a single peak, preceded by a fast rise and followed by a slow decay (see Fig. \ref{fig1}) lasting up to a few minutes. The highly polarized meteor events, assumed to be meteor scatter, typically have quickly varying light curves seldom lasting more than a few seconds. Similarly,  scintillation of radio sources are identified by their irregular variations and random peaks in the light curve over the course of half an hour to few hours  \citep{var19}.

The best differentiation between MRAs and the meteor scatter events would be their spectra, where MRAs are broad band  and scatter events are narrow band. However the spectral information is not saved during typical LASI operation preventing such a test.

\section{Testing the Isotropic Nature of MRAs}
\begin{figure}[h]
 \centering
%
 \includegraphics[width=30pc]{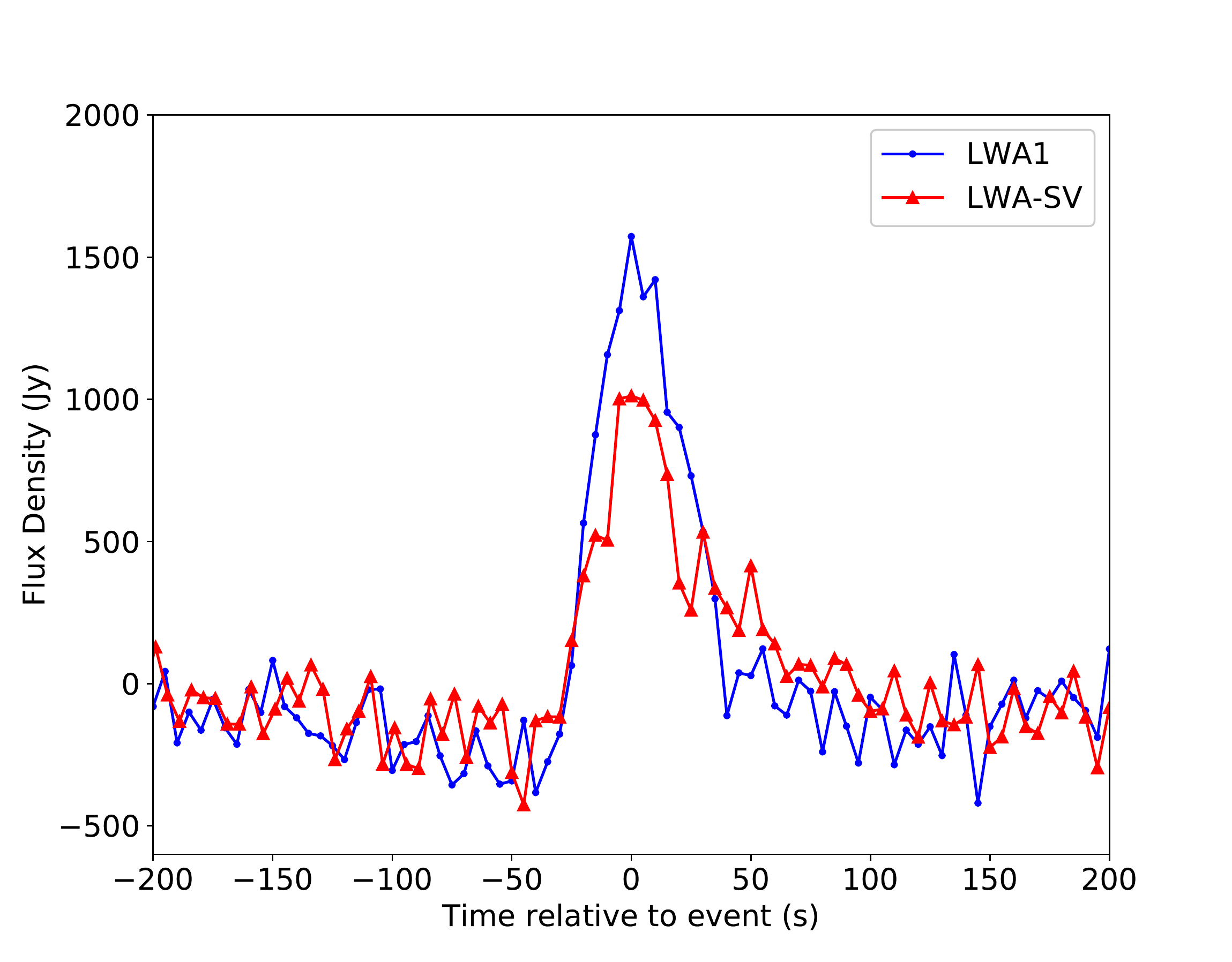}

 \caption{Stokes I light curve of a meteor radio afterglow seen from LWA1 and LWA-SV on MJD 58243 at UTC 15:22:25. The light curve features the fast rise and slow decay of a typical meteor radio afterglow.}
 \label{fig1}
\end{figure}

\begin{figure}[h]
 \centering
%
 \includegraphics[width=30pc]{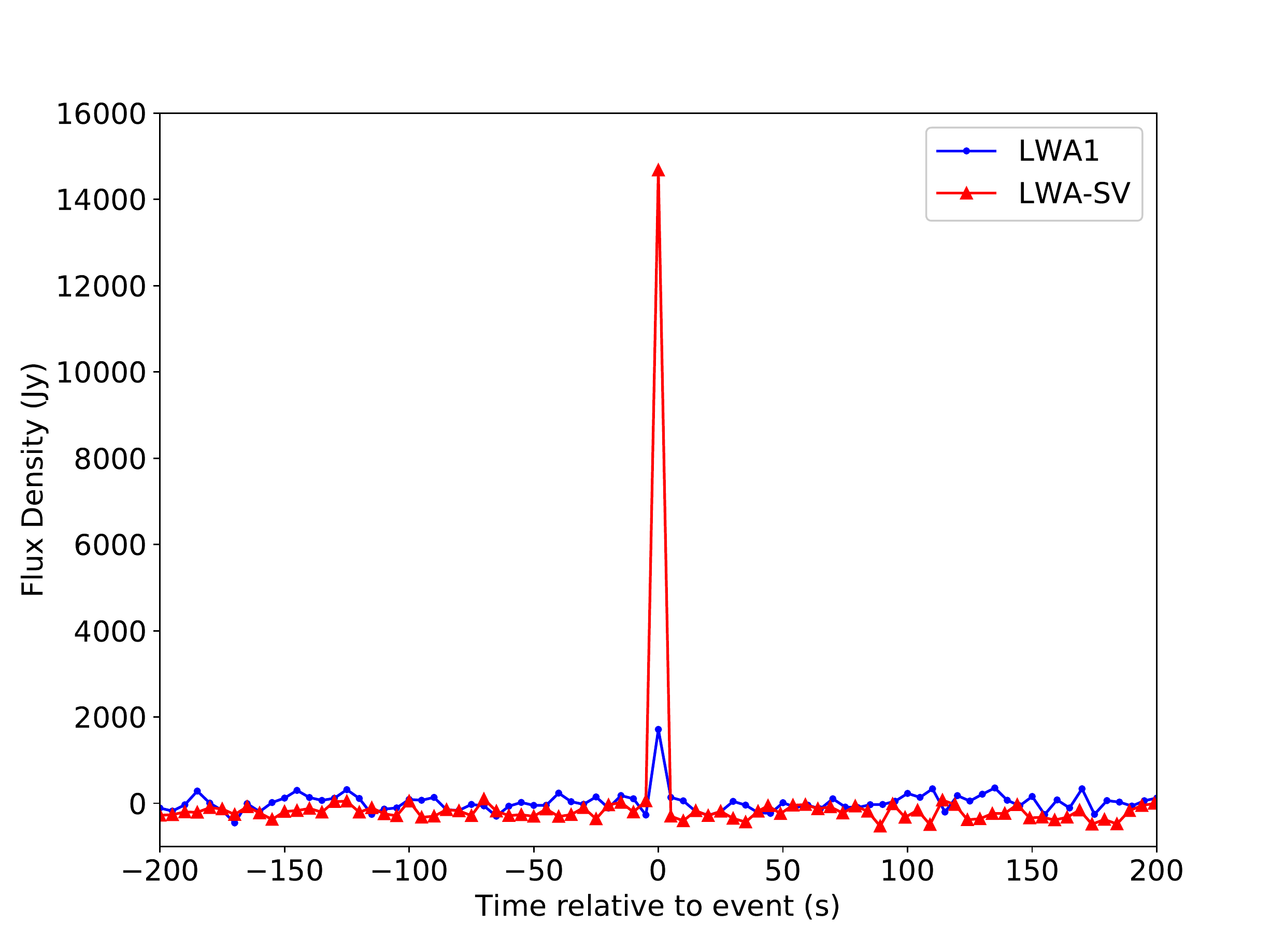}

 \caption{Stokes I light curve of a meteor scatter event seen from LWA1 and LWA-SV on MJD 58172 at UTC 17:12:14. The light curve features the bright, short duration event.}
 \label{fig2}
\end{figure}

Isotropic emitters are sources of EM radiation emitting uniformly in all directions. The flux measured at a distance $r$ away is given by the
 \begin{linenomath*}
  \begin{equation}
F=\frac{L}{4\pi r^{2}},
  \end{equation}
  \end{linenomath*}
 where $F$ is the observed flux and $L$ is the luminosity of source. In this experiment, we are trying to understand the radiation pattern of MRAs by testing whether they are isotropic emitters or not and to compare with known meteor scatter events. Using the two LWA stations, we observed 32 MRAs and 21 meteor scatter events. Fig. \ref{fig1} shows the Stokes I light curve of a co-observed MRA detected in LWA1 and LWA-SV station on MJD 58243. Fig. \ref{fig2} shows the Stokes I light curve of a co-observed meteor scatter event occurred on  MJD 58172.
 
\begin{figure}[h]
 \centering
%
 \includegraphics[width=30pc]{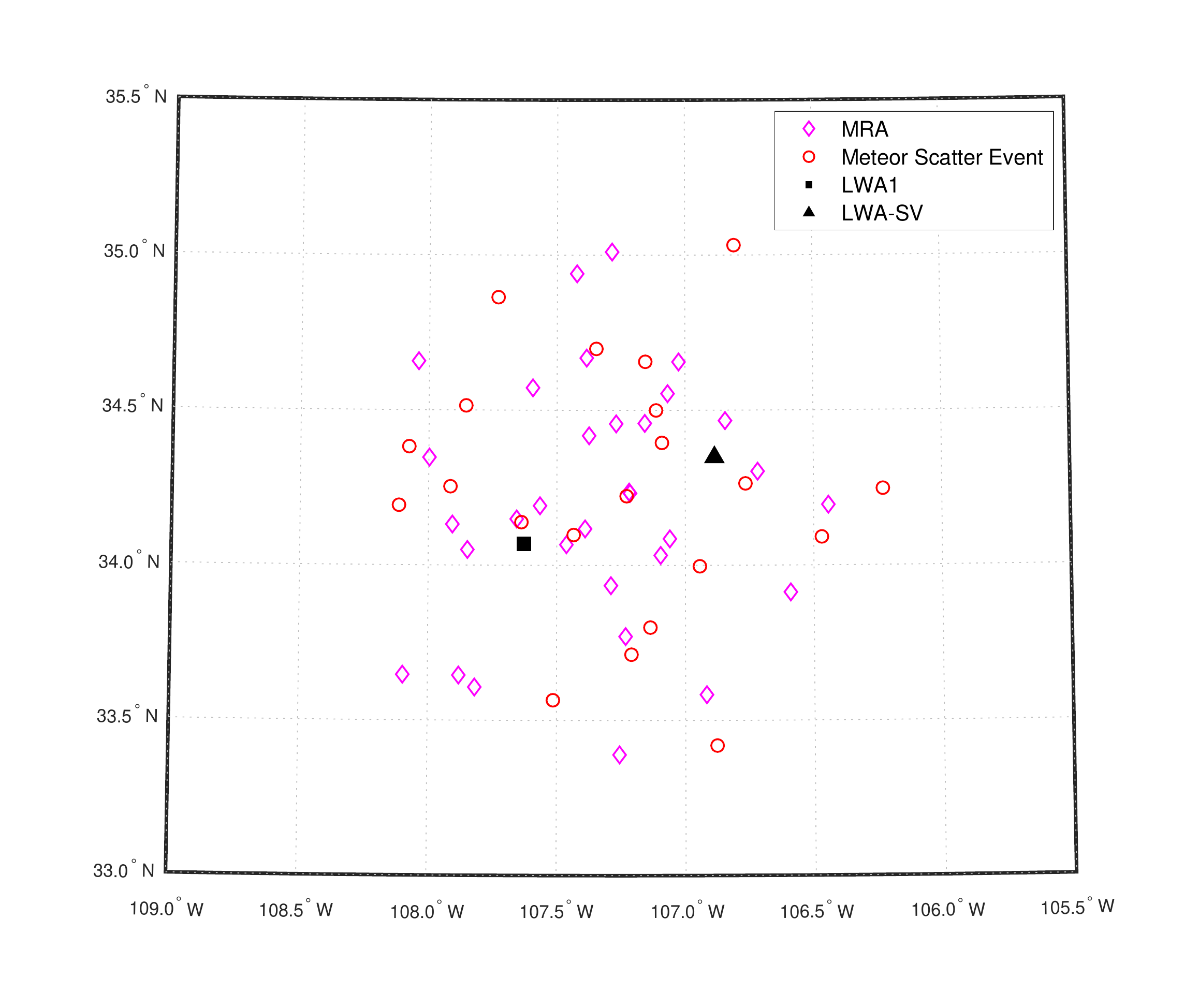}

 \caption{Geographic projection map of 32 MRAs and 21 meteor scatter events collected simultaneously from both stations. The locations of both stations are also marked.}
 \label{fig9}
\end{figure}

%



 The flux of co-observed events can be measured from the all-sky images in each station. From here onwards, we denote $F_{1}$ and $F_{sv}$ to be flux of the transient (MRA/meteor scatter event) measured at the LWA1 and LWA-SV stations respectively. Triangulation gives the physical location of each transient as well as the respective distance to each LWA station. We will denote   $D_{1}$ and $D_{sv}$ to be the distance of transient location from the LWA1 and LWA-SV stations respectively. 
 For an isotropic source, the luminosity measured $L_{1}$ at LWA1 and $L_{SV}$ at LWA-SV have to be equal. Also the flux, $F$, is the product of flux density, $F_{\nu}$, and bandwidth, $\Delta \nu$  (100 kHz in both stations). The flux density in LWA1 and LWA-SV can be denoted as $F_{\nu 1}$ and $F_{\nu SV}$, then we have
\begin{linenomath*}
\begin{equation}
    L_{1} = L_{SV},
\end{equation}
  \begin{equation}
 F_{1}D_{1}^{2} = F_{SV}D_{SV}^{2}, \rm{and}
  \end{equation}
  \begin{equation}
      F_{\nu 1}D_{1}^{2} = F_{\nu SV}D_{SV}^{2}.
  \end{equation}
  \end{linenomath*}
The left hand side and right hand side of [4] can be calculated for different co-observed transients, and the measured luminosity in each station should be equal  if indeed these events are isotropic emitters.

\begin{figure}[h]
 \centering
%
 \includegraphics[width=32pc]{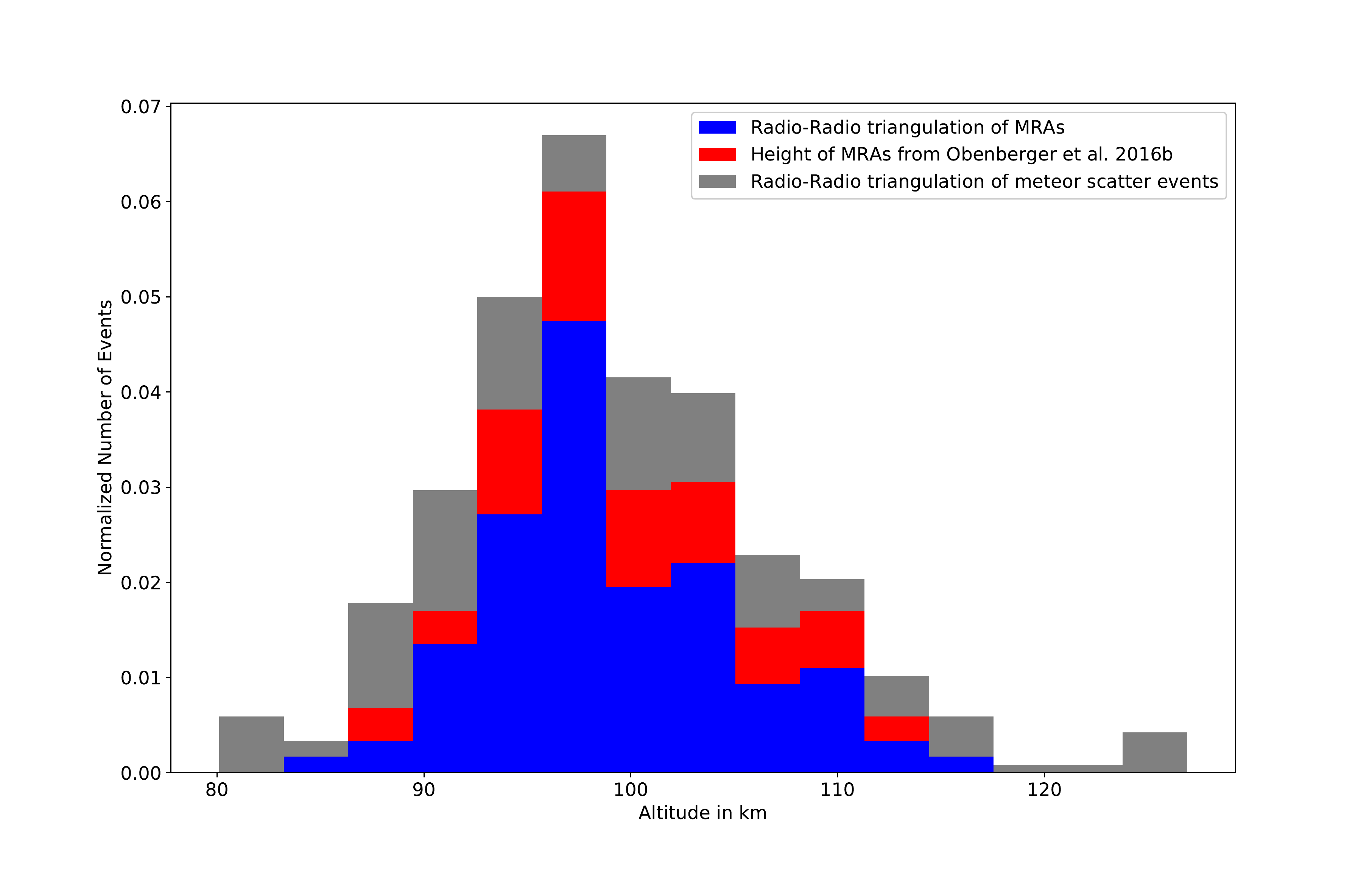}
\caption{Histogram showing the altitude distribution of MRAs and meteor scatter events obtained using different modes of triangulation.}
 \label{fig3}
\end{figure}

Fig. \ref{fig9} shows the geographic projection map of 32 MRAs and 21 meteor scatter events along with the locations of two LWA stations. Both the MRAs and meteor scatter events were detected in a random fashion above both stations and they do not follow any pattern or prefer any occurrence direction above both stations.
Fig. \ref{fig3} shows the altitude distribution of MRAs and meteor scatter events.
The altitudes of 32 MRAs and 21 meteor scatter events have been obtained by radio-radio triangulation using two LWA stations. In addition, the method to determine the MRA altitude from \citet{ob16b} has been used. \citet{ob16b} calculated the altitudes by triangulating the MRAs detected with LWA1 and their optical counterparts detected using two other optical cameras. More details can be found in \cite{ob16b}. The altitude distribution of radio-radio triangulation matches with heights obtained by optical-optical and radio-optical triangulation from  \citet{ob16b}. All three histograms peak between 95-100 km which is the most probable altitude at which MRA and meteor scatter event occurs. This also suggests that the transmitter reflections we observe are indeed from meteor trails. Also, the standard deviation of the meteor scatter distribution is greater than the MRA distribution implying that transmitter reflections can occur at lower as well as higher altitudes as long as a meteor plasma trail is present. We note that the triangulation from meteor scatter events likely contains errors, as the specular reflections seen by each station are most likely coming from different regions of the warped trail.

\subsection{Flux Measurement and Calibration}
The flux is primarily measured using the light curve of the transient. The light curve gives the peak flux and time of emission. For an accurate measurement of flux, we use the all-sky image at the peak time of emission. In this approach, we take the average of 10 noise-like images before the transient and subtract it from the peak flux image. This will remove all the steady sources and the sky residuals from Galactic plane giving the transient detection without losing any flux from it. The next step is to find the peak pixel flux value and thermal noise from a quiet portion of the image in arbitrary units.

\begin{figure}[h]
%
 \includegraphics[width=32pc]{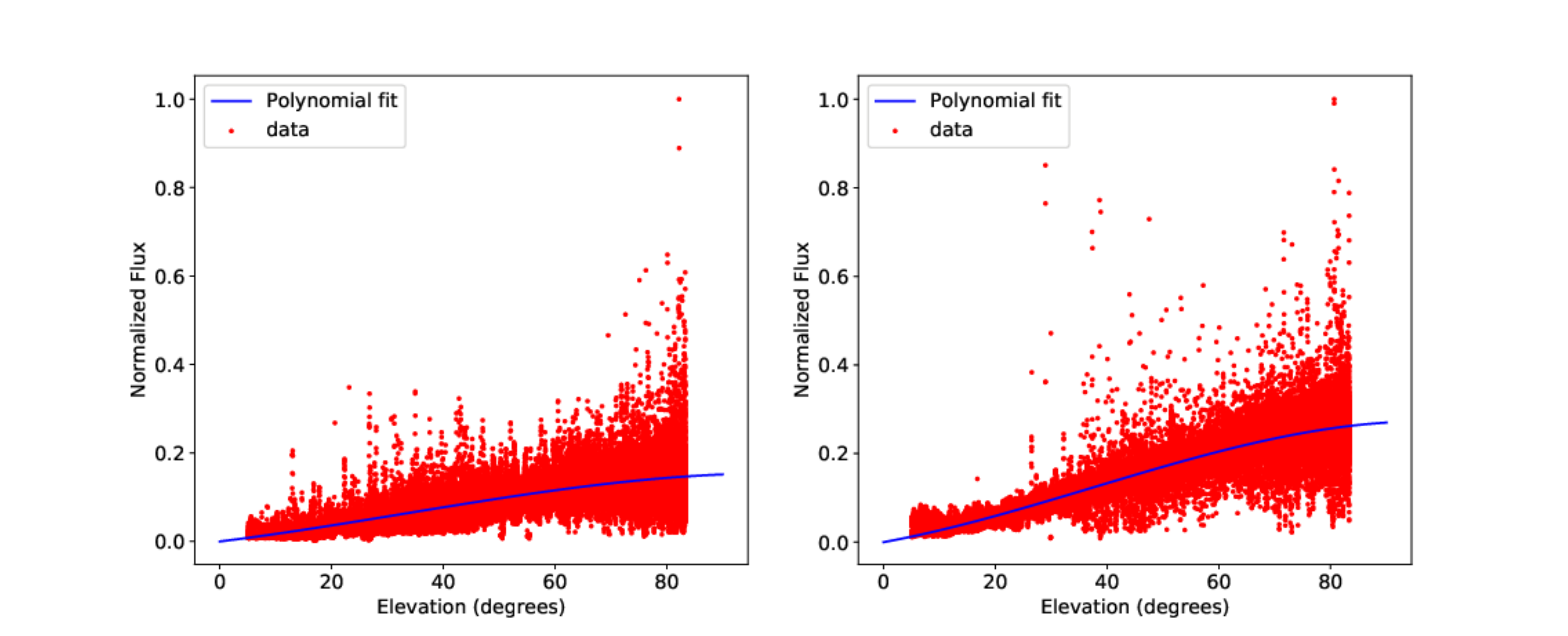}
 \caption{Flux of Cygnus A measured as a function of elevation at 34 MHz (left) and at 38 MHz (right) with LWA 1. The data is fitted with a third order polynomial model.}
 \label{fig4}
\end{figure}

LASI produces uncalibrated images which have pixel values in arbitrary units (a.u.). The pixel values are calibrated using a standard calibrator source in order to convert them to Jansky (1 Jansky (Jy) = $ \rm 10^{-26}\;Watts\;m^{-2}\;Hz^{-1}$). In our case we use Cygnus A, a bright radio galaxy and track the source as it moves across the sky. After tracking the source at each time, the peak flux value of Cygnus A is measured as a function of elevation and frequency. Fig. \ref{fig4} shows the measured flux dependence on frequency and elevation for LWA1.  LASI primarily operates at 34 and 38 MHz and independent flux calibration is needed at each frequency. For a particular frequency and elevation (\textit{elev}),  the scaling relation of the flux density from a.u. to Jy using transient and Cygnus A can be written as
\begin{linenomath*}
  \begin{equation}
\frac{F_{\nu,transient}(elev)\ \rm Jy}{F_{\nu,transient}(elev)\ \rm a.u.}=\frac{F_{\nu,Cyg A}\ \rm Jy}{F_{\nu,Cyg A}(elev)\ \rm a.u.},
  \end{equation}
  \end{linenomath*}
  Rearranging the above equation gives the transient flux density in Jy.
  \begin{linenomath*}
  \begin{equation}
F_{\nu,transient}(elev)\ \rm Jy=\left(\frac{\it F_{\nu,Cyg A}\ \rm Jy}{\it F_{\nu,Cyg A}(elev)\ \rm a.u.}\right)\ \it F_{\nu,transient}(elev)\  \rm a.u., 
  \end{equation}
  \end{linenomath*}
   and the error in measured flux density is given by
  \begin{linenomath*}
  \begin{equation}
\Delta F_{\nu,transient}(elev)\ \rm {Jy}=\left(\frac{\it F_{\nu,Cyg A}\ \rm Jy}{\it F_{\nu,Cyg A}(elev)\ \rm a.u.}\right)\ \it Noise_{\it thermal}\ \rm a.u.  .
  \end{equation}
  \end{linenomath*}
  The calibration was performed separately for each transient event at both stations.
  
\section{Results}
\begin{figure}[h]
 \centering
%
\includegraphics[width=26pc]{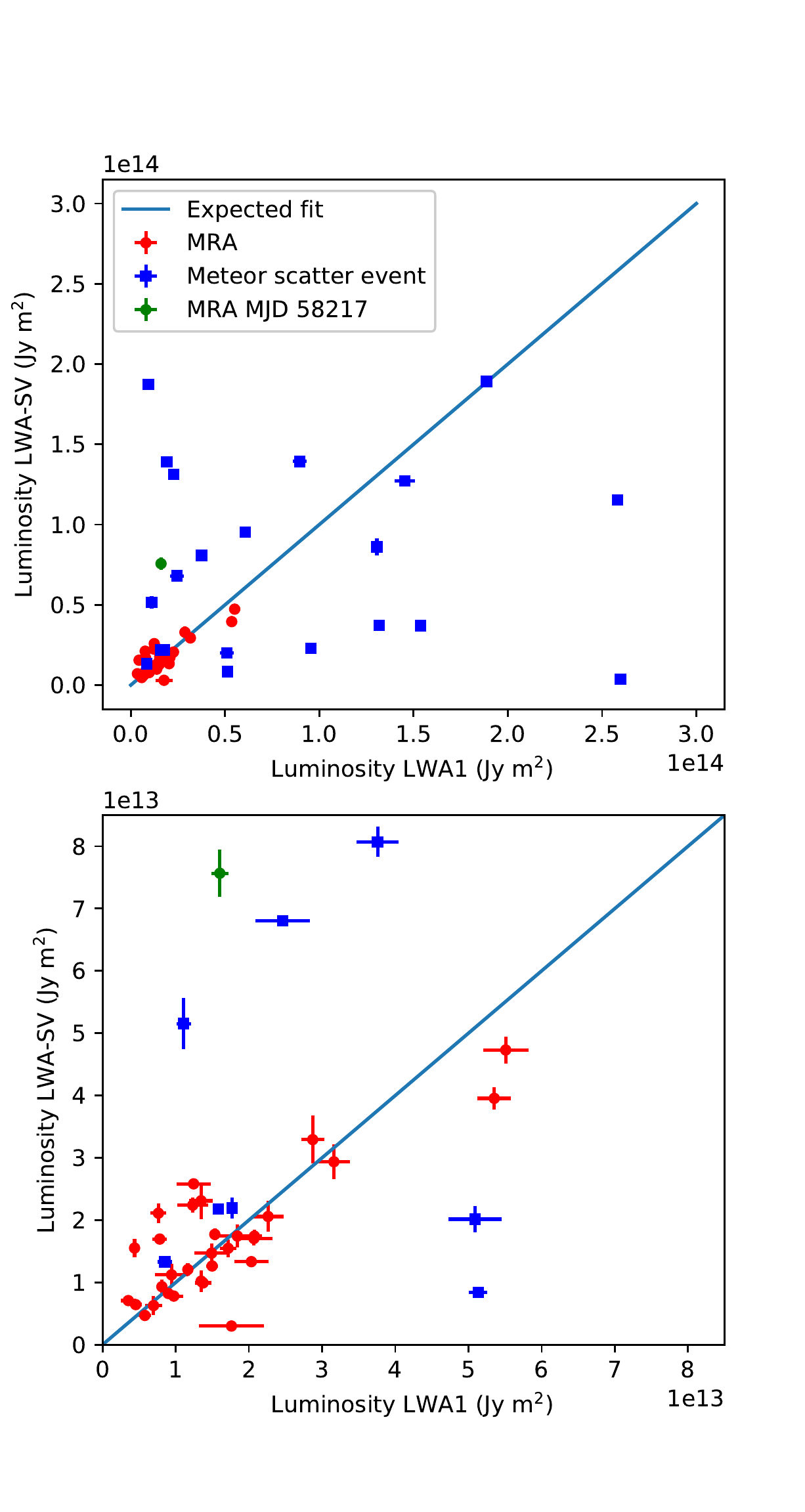}
\caption{The top plot shows the measured luminosity with 1$\sigma$ error bar in each station for 32 MRA events and 21 meteor scatter events. The bottom plot shows the enlarged portion of the region close to the origin in the top plot.  The points marked with red circles are MRAs and with blue squares are meteor scatter events. The calculated p-value of non-correlation for MRAs is $8.47 \times 10^{-9}$  and 0.76 for meteor scatter events.} 
 \label{fig5}
\end{figure}

The distance between the transient location and each LWA station was calculated using the standard distance formula. Theoretical flux densities of Cygnus A at 38 and 34 MHz were calculated using the VLSS Bright Source Spectral Calculator (\url{https://lda10g.alliance.unm.edu/calspec/calspec.html}) where VLSS refers to the VLA Low Frequency Sky Survey at 74 MHz \citep{Coh07,lan12}. The product of flux density and square of distance were calculated for 32 MRAs and 21 meteor scatter events. Fig. \ref{fig5} shows the measured luminosities of MRAs and meteor scatter events from each station.

\begin{figure}[h]
 \centering
%
\includegraphics[width=26pc]{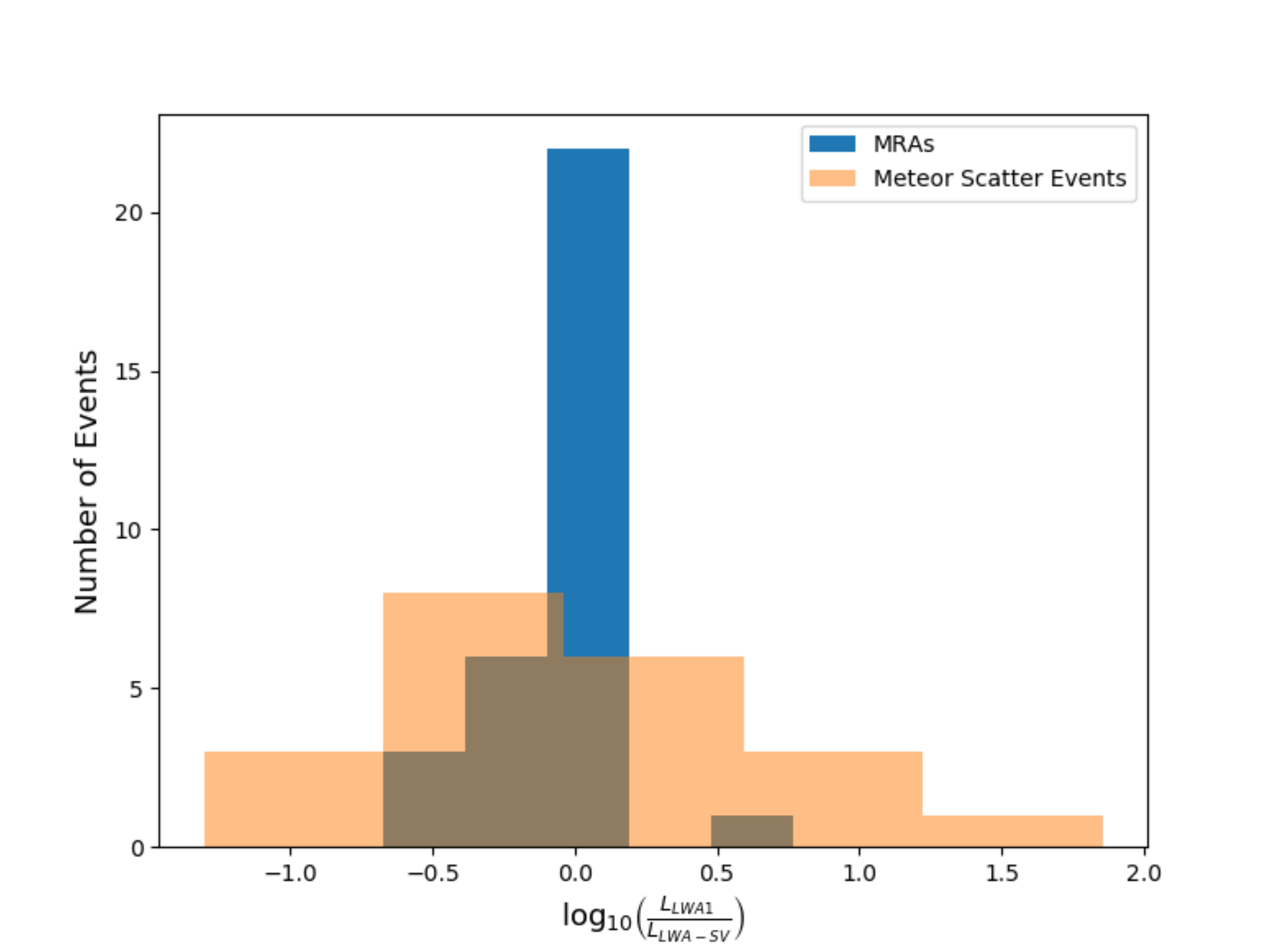}
\caption{Histogram showing the log of ratio of luminosities observed for 32 MRAs and 21 meteor scatter events in two LWA stations.} 
 \label{fig8}
\end{figure}

From the plot it is clear that the MRA events follow the expected linear nature and at the same time the highly polarized events (presumed to be meteor scatter) are randomly scattered around the straight line. However there is an outlier event labeled as MRA MJD 58217 in Fig. \ref{fig5} among the MRAs  which we explain as an elongated MRA. Also Fig. \ref{fig8} shows the histogram distributions of the log of ratio of luminosities measured in each stations. The MRAs have log luminosity ratios peaking at zero with smaller spread and meteor scatter events having wider spread in the data. This suggests that the MRAs have luminosity ratios closer to unity where it deviates far from unity for meteor scatter events.

In statistics, the sample Pearson correlation coefficient measures the correlation within a data sample. A sample Pearson coefficient value of one implies that the sample is highly positively correlated among themselves, zero indicates that the sample is uncorrelated and negative one implies that the sample is highly negatively correlated. Sample Pearson coefficient was calculated for MRAs and meteor scatter events using pearsonr module  (\url{https://docs.scipy.org/doc/scipy/reference/generated/scipy.stats.pearsonr.html}) in the Python Scipy package. This outputs the Pearson correlation coefficient and 2-tailed p-value for non-correlation. A high absolute value of Pearson correlation coefficient and low p-value suggests that data is highly correlated.   The MRAs have a Pearson coefficient of 0.57 and p-value of $6.56 \times 10^{-4}$ with the outlier and Pearson coefficient of 0.83 and p-value of $8.47 \times 10^{-9}$ without the outlier. This suggests that the MRA data has high positive correlation. At the same time, the meteor scatter events have a Pearson coefficient of 0.07 and p-value of 0.76 indicating that the sample is uncorrelated. Also a Chi-square test was carried out to test the goodness of the expected fit with the data.  For MRAs, the calculated $\chi^{2} = 5617$ and for meteor scatter events, $\chi^{2} = 393440$.

\begin{figure}[h]
 \centering
%
 \includegraphics[width=32pc]{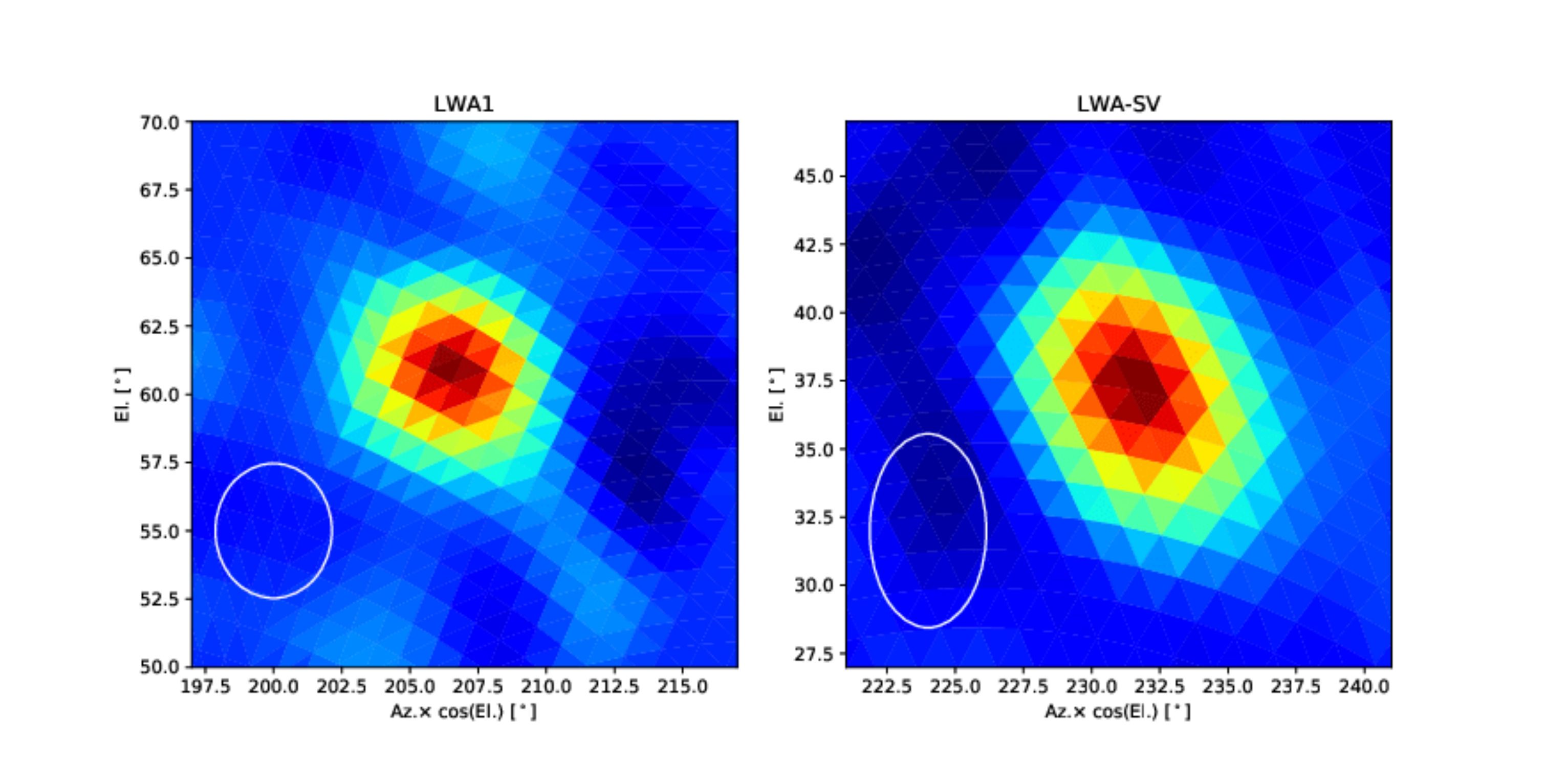}
\caption{Background subtracted image of the MRA event from LWA1 station on MJD 58217 showing the point source structure (left). The image of the MRA event from LWA-SV station on MJD 58217 showing the elongated source structure (right). The ellipse in both images shows the beam shape on the sky at an elevation of 60 degrees in LWA1 and at 37 degrees in LWA-SV where the event occurred. }
 \label{fig6}
\end{figure}

One important reason that may cause a discrepancy between the MRA and the expected linear relationship is projection effects.
The projection effects arise from differences in the orientation of the meteor trail with respect to the line of sight view from each LWA station. If the meteor trail is modeled as a long cylindrical trail, one station can see the projection of the meteor trail as a long rectangle (when the trail is perpendicular to line of sight) and the other station will see the projection as a disk (when the trail is along the line of sight). This results in detecting different integrated surface brightness within a beam area. Fig. \ref{fig6} shows the projection effect of the MRA event on MJD 58217 18:55 UTC from LWA1 and LWA-SV respectively. The azimuth and elevation location at the end points of the source can be used to calculate their angular scale on the sky. The image from LWA1 suggest that that it is a point source structure with a maximum diameter of 9 degrees at an elevation of 60 degrees. At the same time, the LWA-SV image suggest that it has an elongated source structure spanning  approximately 15-16 degrees in length and 8-9 degrees in width across the sky at an elevation of 37 degrees.

%

\section{Discussion}
The emission mechanism responsible for MRAs is currently unknown. Previous work has shown that the emission is non-thermal and for the most part does not have observable polarization. \citet{ob15b} proposed that if electrostatic (Langmuir) plasma waves existed in a meteor trail they could be converted to electromagnetic waves given the steep density gradients within the trail. The fact that the emission spectra follows the expected plasma frequency range of a large meteor trail supports this hypothesis.

A major challenge of this hypothesis, however, is that plasma wave emission is a coherent process, meaning that many electrons oscillate together in unison. Such a processes should then create coherent emission, which would be polarized and highly directional for regions larger than a wavelength ($\sim$ 10 m). Moreover, regions  larger than a Fresnel zone length ($\sqrt{2\lambda R}$  where $\lambda$ is the wavelength and $R$ is the distance to the trail \citep{sug64} i.e. 1400 m at 100 km distance and 10 m wavelength) would be subject to dimming effects associated with nearfield sources. It is certainly possible that many small regions of coherent emission could incoherently add together, collectively producing isotropic radiation without observable polarization.

Alternatively, the emission mechanism may be a completely incoherent process, where a collection of individual randomly oriented electrons radiate in isolation. Such a process would inherently be isotropic and likely not have observable polarization. We note that all of the above mentioned scenarios require a source of energy that provides suprathermal electrons. At this point no source has been identified.

The bottom panel in Fig. \ref{fig5} shows, the measured MRA luminosities deviate between 1\% - 487\%  from the isotropic sphere approximation, however they match this approximation far better than the meteor scatter cases. If we assume MRAs have more or less an even surface brightness then some deviation is expected given that meteor trails are better approximated as thin cylinders rather than spheres. Each station has a unique view of a particular MRA, such that the projected angular size of the cylinder appears different from each station. If we assume a constant surface brightness and optically thick (overdense) medium for plasma trails at low frequencies, the angular size of the projected cylinder would be proportional to the estimated luminosity.

From a randomly generated sample of cylinders with uniform surface brightness, we can estimate what luminosity measurements should look like from each station and then compare to the data presented in this paper. To make this estimate we used a Monte Carlo simulation where we generated 100,000 cylinders with lengths of 10 km and radii of 50 m.  The primary goal of this simulation it is to understand how projection affects the luminosity measurements in each station. We chose 10 km for the length of cylinders as it approximately corresponds to the resolving power of the LWA at 38 MHz at a distance of 100 km. Events that are longer than 10 km can be resolved (i.e. the power is spread over multiple synthesized beams) and they are not included in the simulation. However, if they are shorter than 10 km, they are not resolved and all of flux is measured in one synthesized beam. The generated cylinders are randomly spread over a region $ \pm $ 3 degrees in latitude and $\pm$ 3.6 degrees in longitude from the center of the two arrays. Each meteor has a starting altitude of 100 $\pm$ 10 km and an inclination angle of 45 degrees. Many meteors originate from showers which has a fixed radiant direction and sporadic meteors distribution is not random in the sky \citep{ca09}. Therefore a given meteor will not have a completely random orientation relative to earth. However \citet{hu93} has calculated the most likely meteor incidence angle to be 45 degrees. This will give a better account of the inclination angle without simulating meteors from identical showers and sporadic sources which would be a challenging task.

Since we only search the transient sky above an elevation angle of 30 degrees, we removed all simulated events that had an elevation angle less than 30 degrees in either station. Furthermore, we correct the sample for the elevation angle dependent gain pattern of the LWA antennas \citep{ob15a} and for the fact that the flux of a MRA in a single station drops according the distance to the MRA squared. After applying these filters, roughly 12\% (12,000) of the sample remain, but this sample better represents the actual detected population of MRAs.

\begin{figure}[h]
\centering
%
\includegraphics[width=32pc]{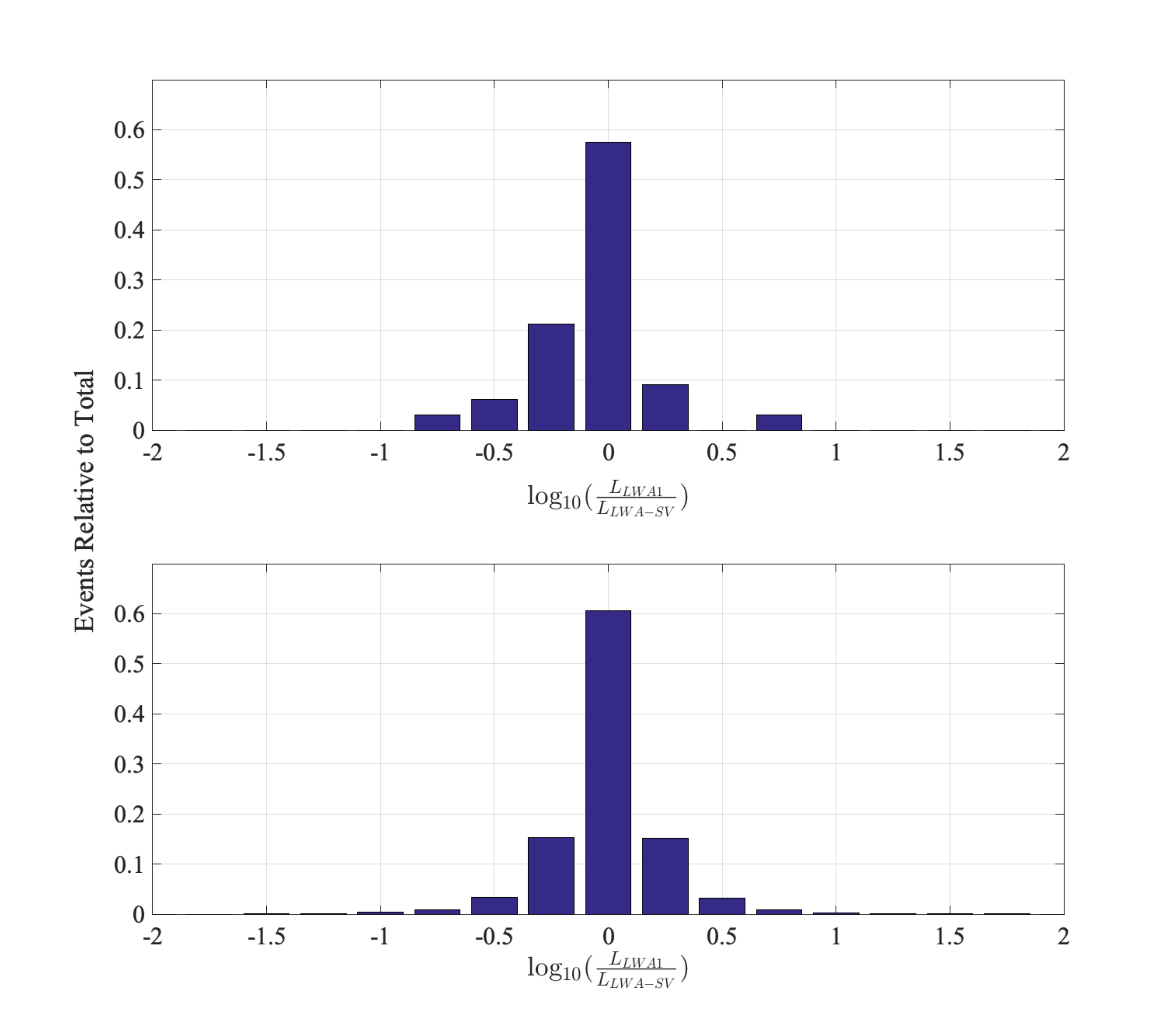}
\caption{Shown are luminosity ratios from LWA1 and LWA-SV for the 32 events presented in this paper (top) and for the Monte Carlo simulation for cylinders of uniform surface brightness (Bottom). The ratios are shown in $\rm{log_{10}}$ base }
 \label{fig7}
\end{figure}

For a given cylinder in this sample, the luminosity estimate from a particular station is simply related to the angular size of projected meteor multiplied by the distance to that station squared. We can then calculate the ratio of the luminosities as measured by each station to find the effect of projection as a deviation from a spherical approximation. Fig. \ref{fig7} shows a histogram of the estimated luminosity ratios from both the simulation and the 32 events presented in this paper. Qualitatively, the two samples are quite similar implying that much of the deviation in the real data can be explained simply by projection effects.

A two sample Kolmogorov-Smirnov (K-S) test can be used to compare the two luminosity ratio distributions in Fig. \ref{fig7}. This method will test if two samples are drawn from the same distribution. The K-S test starts with a null hypothesis that the two samples are drawn from  identical distributions. The two sample K-S test, \url{scipy.stats.ks_2samp} in Python Scipy package will output two values, the K-S statistic and the p-value. 
If the p-value is greater than 10\%, then the null hypothesis that both samples are drawn from the same parent distribution cannot be rejected. For two identical samples, the K-S statistic is 0 and the p-value is 1. Using the luminosity ratio samples from the simulation and from the observed data in Fig. \ref{fig7},  the calculated K-S statistic is 0.15 and p-value is 0.46. This suggests that the samples are similar enough such that the null hypothesis cannot be rejected.

We note that the standard deviation of ratios (in $\rm{log_{10}}$ base) of the 32 events in this paper (0.26) is slightly higher than that of the randomized sample (0.19). While neither sample appears to be Gaussian, the standard deviation gives an idea of the spread in ratios. Randomly selecting 100,000 groups of 32 ratios from the simulated sample shows that there is only a 7\% chance that a group of 32 ratios has a standard deviation greater than 0.26. Therefore, it is likely that our simple projection simulation does not fully explain the variation in luminosity ratios, however it at least suggests that the MRA emission mechanism is not highly directional. This leaves the two possibilities that the emission is either completely incoherent, or it is the result of the incoherent addition of many small coherent emitting regions.  

For these scenarios, sources of variation include uneven surface brightness \citep{ob16b}, irregular shape (not cylindrical) of the radiating region and the presence of neutral winds which can alter the shape of the meteor trails \citep{op09,op14}.  Neither of these situations can readily be simulated, but we should note that the radiating regions are likely not of uniform surface brightness. Indeed, as previous studies \citep{ob16b} have shown, resolved MRAs often have varied brightness across the meteor trail. \citet{op09,op14} developed a new technique to study wind profiles in the lower thermosphere using non-specular meteor echoes. The neutral winds can transport the magnetic field aligned plasma irregularities in the meteor trail. This method has detected intense horizontal wind speeds up to 180 m/s and intense wind shears between 93 and 100 km altitude. The irregularity in the meteor trail shape coming from the neutral winds can cause some deviations for the results obtained in this simulation. Nevertheless this simulation at least explains a large fraction of the luminosity discrepancy between the two stations.

Mentioned above, warping of meteor trails by the neutral wind likely allowed for 21 suspected co-observed specular meteor reflections. These events were identified as being specular meteor scatter using their polarization, light curves and when available, their spectra. Fig. \ref{fig5} shows the radiation pattern of these events are clearly different from the MRAs, therefore we can conclude that MRAs cannot be specular meteor echoes. While non-specular meteor echoes, occurring due to non-field-aligned irregularities, likely scatter in a near isotropic manner, the lack of a bright, specular population, bearing similar characteristics of MRAs (e.g. broadband and unpolarized) suggests that MRAs cannot be non-specular meteors. This evidence strongly supports the previous conclusion that MRAs are indeed emitted by the meteor trails themselves and are not the result of scattered broadband radio frequency signals.

%
\section{Conclusions}
The results suggest that in general MRAs follow an isotropic radiation pattern. In rare cases, the elongation of the afterglow can cause departures from an isotropic radiation pattern. The luminosities only deviate slightly from the spherical approximation. This in turn suggests that the MRAs may be optically thick at our observed frequencies. We find that an isotropically emitting cylinder approximation is a good working model for the MRA emission mechanism. 
The luminosity distribution gives further proof that the detected signals are not reflected sources as they differ greatly from the known meteor scatter events. The isotropic radiation pattern of MRAs also lead us to the conclusion that the emission mechanism could be completely incoherent or the result of incoherent addition of small coherent emitting regions within the plasma trail.  Future high resolution observational studies are required further to clarify the nature of the emission mechanism.

\acknowledgments
Construction of the LWA has been supported by the Office of Naval Research under Contract N00014-07-C-0147 and by the AFOSR. Support for operations and continuing development of the LWA1 is provided by the Air Force Research Laboratory and the National Science Foundation under grants AST-1711164, AST-1835400 and AGS-1708855. The all-sky image data from both LWA stations used in this article are publicly available at the LWA Data Archive
(\url{lda10g.alliance.unm.edu}).

\bibliography{agusample}

\end{document}